\documentclass[review, 3p]{elsarticle}

\usepackage{hyperref}

\journal{Statistics and Probability Letters}
\usepackage{graphicx}
\usepackage{subfigure}
\biboptions{authoryear}

\usepackage{amsmath}

\usepackage{times}
\usepackage{bm}
\usepackage{natbib}
\usepackage{amssymb}
\usepackage{graphicx}

\def\T{{ \mathrm{\scriptscriptstyle T} }}

\newtheorem{thm}{Theorem}
\newtheorem{pp}{Proposition}
\newtheorem{lem}[thm]{Lemma}
\newtheorem{prop}[pp]{Proposition}
\newdefinition{rmk}{Remark}
\newproof{pf}{Proof}
\newproof{pot}{Proof of Theorem \ref{thm2}}

\begin{document}

\begin{frontmatter}

\title{A note on the multiplicative gamma process}

\author{Daniele Durante}
\ead{durante@stat.unipd.it}
\address{Department of Statistical Sciences, University of Padova, Via Cesare Battisti 241, 35121 Padova, Italy}

\begin{abstract}
Adaptive dimensionality reduction in high-dimensional problems is a key topic in statistics. The multiplicative gamma process takes a relevant step in this direction, but improved studies on its properties are required to ease implementation. This note addresses such aim.
\end{abstract}

\begin{keyword}
Matrix factorization \sep Multiplicative gamma process \sep Shrinkage prior \sep Stochastic order.

\end{keyword}

\end{frontmatter}

\section{Introduction}
Increasingly high-dimensional problems have motivated a growing interest towards statistical methodologies which adaptively induce sparsity and automatically delete redundant parameters not required to characterize the data. In these settings it is increasingly useful to leverage representations that have an unknown number of terms. Examples include tensor factorization \citep{dun_2009}, matrix decomposition  \citep{mazu_2010}, mixture modeling  \citep{hjort_2010} and factor analysis  \citep{know_2011}. In addressing related issues, Bayesian methods for high-dimensional problems often rely on purposely overcomplete statistical models, while favoring parsimonious representations via carefully defined priors which facilitate automatic adaptation of the model dimension and lead to efficient estimation for the functionals of interest.

Key examples of such priors include the stick-breaking representation \citep{set_1994}, spike-and-slab priors \citep{ish_2005} and the multiplicative gamma process \citep{bha_2011} --- among others.  These proposals facilitate shrinkage by  defining priors that are potentially able to concentrate increasing mass around values deleting the effect of parameters associated with growing dimensions of the statistical model. For example, the stick-breaking prior for the weights in a mixture model assigns growing mass towards zero for the weights associated with an increasing number of mixture components --- for appropriate hyperparameter settings. In such cases, as the number of components grows, their weights tend to concentrate close to zero a priori, reducing the importance of the later components in defining the mixture representation. 

Although these procedures provide strategies to deal with high-dimensional data, their performance  may be sensitive to several choices including model specification, hyperparameter settings and prior choices for other quantities not directly related to shrinkage;  refer to \citet{roos_2011}  for a discussion on similar issues.  Hence,   in making these procedures standard practice, it is important to provide researchers with strategies to check whether the prior actually achieves the shrinkage they seek when using these methods. In fact, a naive choice of hyperparameters may lead to a lack of ability to avoid overfitting and adaptively remove the unnecessary components from the model.

This note aims to provide an improved characterization for the probabilistic properties of the multiplicative gamma process prior recently developed by  \citet{bha_2011} for Bayesian inference on  large covariance matrices. Their approach has proven useful in quite broad settings involving factorizations having an unknown number of terms.  This includes not only the original high-dimensional factor model motivation, but also other matrix factorization and basis expansion settings.  For some examples of recent applications refer to \citet{mont_2012}, \citet{zhu_2014}, \citet{rai_2014} and \citet{lian_2014} --- among others. Given the popularity of the multiplicative gamma process shrinkage approach, it is important to carefully study the probabilistic properties of the prior, including the role of the hyperparameters in controlling shrinkage.

 \section{The multiplicative gamma process and motivations}
  \label{sec1}
Let $y_{i}=(y_{i1}, \ldots, y_{ip})^\T \in \Re^{p}$,  $i=1, \ldots, n$, denote independent and identically distributed observations from a $p$-variate Gaussian distribution $\mbox{N}_p(0, \Omega)$ with $p \times p$ covariance matrix $\Omega$. The goal of  \citet{bha_2011} is to provide Bayesian inference on the dependence structure among the $p$ variables under study. 

This aim is accomplished by modeling the covariance matrix $\Omega$ via matrix factorization representations, which enhance borrowing of information and facilitate adaptive dimensionality reduction via careful prior specification. Hence, instead of direct modeling of elements $\omega_{js}$ comprising the matrix $\Omega$, they consider a parameter expanded representation $\Omega=\Lambda \Lambda^\T + \Sigma$, which arises by marginalizing out the latent factors $\eta_i \sim \mbox{N}_k(0, \mbox{I}_k)$, $i=1, \ldots, n$, in the model $y_i = \Lambda \eta_i + \epsilon_i$, where $\Lambda$ is a $p \times k$ matrix of factor loadings and $\epsilon_i$, $i=1, \ldots, n$, are independent and identically distributed noise vectors from a $\mbox{N}_p(0, \Sigma)$  with $p \times p$ covariance matrix  $\Sigma=\mbox{diag}(\sigma^2_1, \ldots, \sigma_p^2)$. 

As the identifiability of the factor loadings is not required for the estimation of the covariance matrix $\Omega$,  \citet{bha_2011} avoid model constraints on $\Lambda$, but instead leverage their parameter expanded representation which is purposely over-parameterized to obtain improved computational strategies via simple Markov chain Monte Carlo methods and useful  properties including model flexibility, prior support and posterior consistency.

\begin{table}[t]
\begin{center}
\def~{\hphantom{0}}
{%
\begin{tabular}{lcccccccccccccccc}
& &&& \multicolumn{3}{c}{$\tau_{h}$ }  & && & \multicolumn{3}{c}{$\theta_h=1/\tau_{h}$  }&&&&\multicolumn{1}{c}{$\lambda_{jh}$ }\\ 
\hline 
&&&& $Q_1$ & $Q_2$  & $Q_3$  & && &$Q_1$ &  $Q_2$&  $Q_3$ &&& &$Q_3-Q_1$\\
\hline
 ~$h=1$ &&&& $0{.}29$ & $0{.}69$ & $1{.}39$  &&&& $0{.}72$& $1{.}44$&  $~3{.}50$&&&& $1{.}63$\\
 ~$h=2$ &&&& $0{.}13$ & $0{.}45$ & $1{.}28$&&&& $0{.}78$& $2{.}20$&  $~7{.}50$&&&& $1{.}91$\\
 ~$h=3$ &&&& $0{.}07$ & $0{.}30$ & $1{.}08$ &&&& $0{.}92$& $3{.}35$&  $14{.}86$&&&& $2{.}29$\\
 ~$h=4$ &&&& $0{.}04$ & $0{.}19$ & $0{.}88$ &&&& $1{.}13$& $5{.}12$&  $28{.}16$&&&& $2{.}77$\\
 \hline
\end{tabular}}
\caption{Stochastic behavior of the priors for the model parameters. For $a_1=1$ and $a_2=1{.}1>1$, first ($Q_1$), second ($Q_2$) and third ($Q_3$) quartiles of priors \eqref{eq3}--\eqref{eq4} for the column-specific $\tau_h$ and $\theta_h=1/\tau_h$ at increasing column index $h=1, \ldots, 4$. The last column displays the interquartile range for the prior induced on the factor loadings $\lambda_{jh}$ via \eqref{eq2} with $\tau_h$ from  \eqref{eq3}. Without loss of generality, local parameters $\phi_{jh}$ are set at $1$. Quantities are obtained from $1{,}000{,}000$ samples from the priors  \eqref{eq3}--\eqref{eq4} and  \eqref{eq2}, respectively.}
\label{table1}
\end{center}
\end{table}

Potentially an over-parameterized likelihood is associated with an increased risk of overfitting the model with respect to the observed data, unless careful shrinkage procedures are introduced to penalize the introduction of too many factors in the covariance matrix factorization. To avoid poor performance in inference and prediction, \citet{bha_2011} enhance adaptive dimensionality reduction by defining prior distributions  for the loadings $\lambda_{jh}$, $j=1, \ldots, p$ and $h=1, \ldots, k$, which are carefully designed to penalize a priori the contribution of increasing columns of $\Lambda$ in defining $\Omega$, via $\Omega=\Lambda \Lambda^\T + \Sigma$. Specifically they let $\sigma^{-2}_j \sim \mbox{Ga}(a_{\sigma},b_{\sigma})$, for each $j=1, \ldots, p$, and propose the following  multiplicative gamma process prior which aims to increase the degree of shrinkage as the column index of $\Lambda$ grows,
\begin{eqnarray}
\lambda_{jh} \mid \phi_{jh}, \tau_h \sim {\mbox N}(0, \phi_{jh}^{-1} \tau_h^{-1}), \quad \phi_{jh} \sim \mbox{Ga}(\upsilon/2, \upsilon/2), \quad j=1, \ldots, p, \ \  h=1, \ldots, k,
\label{eq2}
\end{eqnarray}
with the column-specific precision parameters $\tau_{h}$ defined via a cumulative product of independent gamma distributions
\begin{eqnarray}
\tau_{h}=\prod_{l=1}^h \delta_l, \quad \delta_1 \sim \mbox{Ga}(a_1,1), \ \ \ \delta_{l\geq 2} \sim \mbox{Ga}(a_2,1),  \quad h=1, \ldots, k.
\label{eq3}
\end{eqnarray}
Hence in \eqref{eq2}, quantities $ \phi_{jh}$ measure local precision of each loading element, while $\tau_h$ is a column-specific parameter which controls the global shrinkage for the $h$th column of $\Lambda$. According to \eqref{eq2}, when $1/\tau_{h}\approx 0$, the priors for the loadings in the $h$th column  $\lambda_h=(\lambda_{1h}, \ldots, \lambda_{ph})^\T$  concentrate considerable mass  around zero, meaning that the $h$th column has minimal effect a priori in defining the covariance matrix $\Omega$ via $\Omega=\Lambda \Lambda^\T + \Sigma$. According to the characterization of $\tau_h$ in  \eqref{eq3} this shrinkage property is not constant across the columns of $\Lambda$, but is designed to be potentially stronger as the column index increases in order to penalize growing dimensional factorizations of $\Omega$.

The type of shrinkage induced by the multiplicative gamma prior in \eqref{eq3} depends on the values of the hyperparameters $a_1>0$ and $a_2>0$. Although \citet{bha_2011} consider hyperpriors on $a_1$ and $a_2$ to learn these quantities from the data, a wide set of methods building on their contribution fix such hyperparameters following some of the authors' statements on the behavior of  \eqref{eq2}--\eqref{eq3} in relation to $a_1$ and $a_2$. In particular,  \citet{bha_2011}  claim that the quantities $\tau_h$ are stochastically increasing under the restriction $a_2>1$, meaning that the  induced prior on $1/\tau_h$ increasingly shrinks  towards zero as the column index $h$ grows,  when $a_2>1$. Although it is true that $\mbox{E}(\tau_h)=\mbox{E}(\prod_{l=1}^h \delta_l)=\prod_{l=1}^h\mbox{E}(\delta_l)=a_1a_2^{h-1}$ increases with $h$ when $a_2>1$, such result is not sufficient to guarantee the increasing shrinkage property motivating the multiplicative gamma process construction. 

As shown in Table \ref{table1}, a value of $a_2=1{.}1>1$ induces priors on parameters $\tau_h$ which seem stochastically decreasing as $h$ increases. This leads to apparently stochastically increasing distributions on $1/\tau_h$, which in turn enforce priors on the loadings $\lambda_{jh}$ that are increasingly diffuse as the column index $h$ increases. Hence, researchers choosing $a_2=1{.}1$ will obtain a prior with an apparently opposite behavior with respect to the one they seek when using a multiplicative gamma process. Beside this, even increasing expectation in $\tau_h$ does not necessarily imply  decreasing expectation in  $1/\tau_h$, and therefore growing shrinkage on average. In fact,  the multiplicative gamma prior \eqref{eq3} on $\tau_h$ induces a multiplicative inverse gamma prior on the column-specific variance parameter $\theta_h=1/\tau_h$, which is defined as 
\begin{eqnarray}
\theta_h=\frac{1}{\tau_h}=\frac{1}{\prod_{l=1}^h \delta_l}=\prod_{l=1}^h \frac{1}{ \delta_l}=\prod_{l=1}^h \vartheta_l, \quad \vartheta_1 \sim \mbox{Inv-Ga}(a_1,1), \ \ \ \vartheta_{l \geq 2} \sim \mbox{Inv-Ga}(a_2,1),  \quad h=1, \ldots, k.
\label{eq4}
\end{eqnarray}
As a result $\mbox{E}(\theta_h)=\mbox{E}(\prod_{l=1}^h \vartheta_l)=\prod_{l=1}^h\mbox{E}(\vartheta_l)=1/\{(a_1-1)(a_2-1)^{h-1}\}$. Hence for values $1<a_2<2$ both $1/\tau_h$ and $\tau_h$ increase across the columns of $\Lambda$ in expectation. 

Motivated by these misleading results, this note aims to improve the probabilistic characterization of the multiplicative gamma process and add further insights on the properties of this prior compared to those  available in  \citet{bha_2011}, including simple strategies to check whether specific hyperparameter choices actually facilitate shrinkage. Although it is possible to define other priors characterized by shrinkage properties, there are two additional features associated with the multiplicative gamma process, making this prior a powerful candidate in several applications and, therefore, worth a deeper probabilistic understanding. Firstly, this prior is characterized by simple strategies for posterior computation relying on conjugate full conditionals updates. Moreover, as outlined in Proposition \ref{prop0}, the multiplicative inverse gamma prior in \eqref{eq4} has full support on the positive $k$-dimensional Euclidean space $(0, +\infty)^k$. This a key property to enhance flexibility in the type of shrinkage induced on the factor loadings.
\begin{prop}
Let $(\theta_1, \ldots, \theta_k)$ denote a random vector distributed according to the multiplicative inverse gamma prior in equation \eqref{eq4}. Then, for any $(\theta^0_1, \ldots, \theta^0_k) \in  (0, +\infty)^k$ and $k$, $\mbox{\normalfont pr}(\sum_{h=1}^k |\theta_h-\theta^0_h |< \varepsilon)>0$, for every $\varepsilon>0$.
\label{prop0}
\end{prop}
\begin{pf}
To prove the above Proposition, note that a lower bound for $\mbox{\normalfont pr}(\sum_{h=1}^k |\theta_h-\theta^0_h |< \varepsilon)$  is given by $\mbox{\normalfont pr}(|\theta_h-\theta^0_h |< \varepsilon/k, \mbox{ for all } h=1, \ldots, k)$. Letting $\mathbb{B}_{\varepsilon}(\Theta^0)=\{ (\theta_1, \ldots, \theta_k): |\theta_h-\theta^0_h |< \varepsilon/k, \mbox{ for all } h=1, \ldots, k\}$ and
exploiting the representation of the multiplicative inverse gamma prior in equation \eqref{eq4}, the lower bound probability can be expressed as $\int_{\mathbb{B}_{\varepsilon}(\Theta_0)} f_{\theta_1}(\theta_1) \prod_{h=2}^k f_{\theta_h \mid \theta_{h-1}}(\theta_h) {\mbox d} (\theta_1, \ldots, \theta_k)$, where $f_{\theta_h \mid \theta_{h-1}}(\theta_h)$ is the conditional density function of $\theta_h$ given $\theta_{h-1}$. Hence, the joint prior for  $(\theta_1, \ldots, \theta_k)$  can be factorized as the product of conditional densities with $\theta_1=\vartheta_1   \sim \mbox{Inv-Ga}(a_{1},1)$ and $\theta_{h} \mid  \theta_{h-1}= \theta_{h-1}\vartheta_h\sim \mbox{Inv-Ga}(a_{2}, \theta_{h-1})$, for each $h=2, \ldots, k$. Therefore, since the $ \mbox{Inv-Ga}(a,b)$ has full support in $(0, +\infty)$ for any $a>0, b>0$ and provided that by definition $ \theta_{h-1}>0$ for each $h=2, \ldots, k$, it follows that $\mbox{\normalfont pr}(|\theta_h-\theta^0_h |< \varepsilon/k, \mbox{ for all } h=1, \ldots, k)>0$ for any $k$ and $ \varepsilon>0$. \qed
\end{pf}

 \section{Stochastic order and shrinkage in the multiplicative inverse gamma prior}
   \label{sec2}
Consistent with the discussion in Section \ref{sec1}, let us study the stochastic behavior of the sequence $\theta_1, \ldots, \theta_k$, with each $\theta_h$ defined as the product of $h$ independent inverse gamma random variables $\vartheta_1, \ldots, \vartheta_h$, for any $h=1, \ldots, k$. Stochastic order $\theta_1\succeq \cdots \succeq \theta_k$ --- if holding --- would be an appealing property for the multiplicative inverse gamma prior in guaranteeing --- by definition --- that the mass $\mbox{pr}\{\theta_{h}\in (0,\theta]\}=\mbox{pr}(\theta_{h}\leq \theta)$ assigned by the prior  to the interval $(0,\theta]$, increases with $h$, for every $\theta >0$. This facilitates increasing shrinkage across the columns of $\Lambda$. Exploiting the transitivity property of the stochastic order \citep[e.g.][page 30]{bel_2015}, this requires showing $\mbox{pr}(\theta_{h+1}\leq \theta)\geq \mbox{pr}(\theta_{h}\leq \theta)$ for any $h=1, \ldots, k-1$ and $\theta >0$, or equivalently that $ {\mbox{E}}\{\psi( \theta_{h+1})\}\leq  {\mbox{E}}\{\psi( \theta_{h})\}$ for all the increasing functions $\psi(\cdot)$, for which expectation exists  \citep[e.g.][page 4]{shak_2007}.  Lemma \ref{teo1} states that the condition $a_2>1$, suggested by \citet{bha_2011}, does not ensure stochastic order. 
\begin{lem}
There exist infinitely many multiplicative inverse gamma priors with $a_2>1$, such that $\theta_1\nsucceq \cdots \nsucceq \theta_k$.
\label{teo1}
\end{lem}

\begin{pf}
In proving Lemma \ref{teo1}, let us first derive the quantity $\mbox{E}(\theta_h^m)$, with $m>0$. According to \eqref{eq4}, this requires first $\mbox{E}(\vartheta^m)$, where $\vartheta=1/\delta$ is a generic inverse gamma random variable $\vartheta \sim \mbox{Inv-Ga}(a,1)$. Hence
\begin{eqnarray}
\mbox{E}(\vartheta^m)&=& \int_{0}^{+\infty}\vartheta^m \frac{1}{\Gamma(a)} \vartheta^{-a-1}e^{-1/ \vartheta} {\rm d} \vartheta=\frac{1}{\Gamma(a)}\int_{0}^{+\infty}\vartheta^{-(a-m)-1}e^{-1/ \vartheta} {\rm d} \vartheta, \nonumber \\
& =&\frac{\Gamma(a-m)}{\Gamma(a)}\int_{0}^{+\infty}\frac{1}{\Gamma(a-m)}\vartheta^{-(a-m)-1}e^{-1/ \vartheta} {\rm d} \vartheta=\frac{\Gamma(a-m)}{\Gamma(a)}, \quad 0<m<a.\ \ \  \ \ 
\label{invga}
\end{eqnarray}
Exploiting \eqref{invga}, $\mbox{E}(\theta_h^m)=\mbox{E}\{(\prod_{l=1}^h \vartheta_l )^m\}=\mbox{E}(\prod_{l=1}^h \vartheta_l^m)=\{\Gamma(a_1-m)/\Gamma(a_1)\}\{\Gamma(a_2-m)^{h-1}/\Gamma(a_2)^{h-1}\}$, $0<m<a_1$, $0<m<a_2$. Based on this result, the proof proceeds by showing that there always exists an increasing function $\psi^*(\cdot)$ for which $\mbox{E}\{\psi^*(\theta_{h})\}$ exists and such that  $\mbox{E}\{\psi^*(\theta_{h+1})\} > \mbox{E}\{\psi^*(\theta_{h})\}$, for every $h=1, \ldots, k-1$. 

Let $a_1\geq a_2$ and $0<m_{a_2}<a_2$, and consider $\psi^*(\theta_h)=\theta_h^{m_{a_2}}$. As $m_{a_2}$ is positive and since  $\theta_h \in (0,+\infty)$ for every $h=1, \ldots, k$, the function $\psi^*(\theta_h)$ is increasing in $ (0,+\infty)$. Moreover, as $m_{a_2}<a_2\leq a_1$,  $\mbox{E}\{\psi^*(\theta_{h})\}$ exists for every $h=1, \ldots, k$. Exploiting results in \eqref{invga},
\begin{eqnarray}
 \mbox{E}(\theta_{h+1}^{m_{a_2}})- \mbox{E}(\theta_{h}^{m_{a_2}})&=&\frac{\Gamma(a_1-m_{a_2})}{\Gamma(a_1)}\frac{\Gamma(a_2-m_{a_2})^{h}}{\Gamma(a_2)^{h}}-\frac{\Gamma(a_1-m_{a_2})}{\Gamma(a_1)}\frac{\Gamma(a_2-m_{a_2})^{h-1}}{\Gamma(a_2)^{h-1}} \nonumber\\
 &=&\frac{\Gamma(a_1-m_{a_2})}{\Gamma(a_1)}\frac{\Gamma(a_2-m_{a_2})^{h-1}}{\Gamma(a_2)^{h-1}} \left\{ \frac{\Gamma(a_2-m_{a_2})}{\Gamma(a_2)}-1\right\}.
 \label{prof}
\end{eqnarray}
Hence showing $ \mbox{E}(\theta_{h+1}^{m_{a_2}})- \mbox{E}(\theta_{h}^{m_{a_2}})>0$, requires finding a value of $m_{a_2}$ such that $0<m_{a_2}<a_2$ and $\Gamma(a_2-m_{a_2})>\Gamma(a_2)$. According to the well known properties and functional form of the gamma function $\Gamma(\cdot)$, it is always possible to find a value $m_{a_2}$ less than $a_2$ but sufficiently close to $a_2$, such that their difference $a_2-m_{a_2}$ is close enough to zero to obtain $\Gamma(a_2-m_{a_2})>\Gamma(a_2)$. This proves Lemma \ref{teo1}. \qed
\end{pf}

Although absence of stochastic order in the entire sample space $(0,+\infty)$ is an undesired property, it does not necessarily rule out the cumulative shrinkage behavior for which the prior has been developed. In fact,  order in probabilities $\mbox{pr}\{\theta_1 \in (0, \theta] \} \leq \cdots \leq \mbox{pr}\{\theta_h \in (0, \theta] \}\leq  \mbox{pr}\{\theta_{h+1} \in (0, \theta] \} \leq \cdots \leq \mbox{pr}\{\theta_k \in (0, \theta] \}$ for reasonable neighborhoods $(0, \theta]$ of zero --- instead of stochastic order in  the entire sample space $(0,+\infty)$ ---  can be sufficient to facilitate shrinkage. According to Lemma \ref{teo2} this property holds for every $a_1>0$ and $a_2>0$ as $\theta \rightarrow 0^{+}$. 

\begin{lem}
For all $a_1>0$ and $a_2>0$, $\lim_{\theta \to 0^{+}}[{ \normalfont \mbox{pr}}\{\theta_{h+1} \in (0, \theta] \}/{ \normalfont \mbox{pr}}\{\theta_{h} \in (0, \theta] \}]\geq 1$, for any $h=1, \ldots, k-1$.
\label{teo2}
\end{lem}

\begin{pf}
As ${ \normalfont \mbox{pr}}\{\theta_{h+1} \in (0, \theta] \}/{ \normalfont \mbox{pr}}\{\theta_{h} \in (0, \theta] \}=F_{\theta_{h+1}}(\theta)/F_{\theta_{h}}(\theta)$ is the ratio of the  cumulative distribution functions of  $\theta_{h+1}$ and $\theta_{h}$, respectively, and provided that their corresponding probability density functions  $f_{\theta_{h+1}}(\theta)$ and $f_{\theta_{h}}(\theta)$ are positive in $(0, +\infty)$, it suffices to show that $f_{\theta_{h+1}}(\theta)$  is an upper bound of $f_{\theta_{h}}(\theta)$, when $\theta \to 0^{+}$. Therefore --- since both $f_{\theta_{h+1}}(\theta)$ and $f_{\theta_{h}}(\theta)$ are positive functions in $(0, +\infty)$ ---  Lemma \ref{teo2} holds if  $\lim_{\theta \to 0^{+}} \{f_{\theta_{h+1}}(\theta)/f_{\theta_{h}}(\theta) \}\geq 1$ for every $a_1>0$, $a_2>0$ and $h=1, \ldots, k-1$.

Let us first consider the limit inequalities for $h=2, \ldots, k-1$. As the probability density function for a product of independent gamma random variables is available via sophisticated  Meijer G-functions \citep{spri_1970}, let us first focus on proving  $\lim_{\theta \to 0^{+}} \{f_{\theta_{h+1} \mid \theta_{h-1}}(\theta)/f_{\theta_{h} \mid \theta_{h-1}}(\theta) \}\geq 1$, for every $h=2, \ldots, k-1$, with  $f_{\theta_{h+1} \mid \theta_{h-1}}(\theta)$ and $f_{\theta_{h} \mid \theta_{h-1}}(\theta)$ the conditional density functions of $\theta_{h+1}$ and $\theta_{h}$, given $\theta_{h-1}>0$, respectively. As $\theta_{h}=\theta_{h-1}\vartheta_h$, with $\vartheta_h \sim \mbox{Inv-Ga}(a_2,1)$, from the standard properties of the inverse gamma random variable, $f_{\theta_{h} \mid \theta_{h-1}}(\theta)$ is easily available as the probability density function for the random variable $\theta_{h} \mid \theta_{h-1} \sim  \mbox{Inv-Ga}(a_2,\theta_{h-1})$. To compute $f_{\theta_{h+1} \mid \theta_{h-1}}(\theta)$ note instead that $\theta_{h+1} =\theta_{h}\vartheta_{h+1}$, with $\vartheta_{h+1} \sim  \mbox{Inv-Ga}(a_2,1) $. Hence $\theta_{h+1} \mid \theta_{h}\sim  \mbox{Inv-Ga}(a_2,\theta_h)$ and $\theta_{h} \mid \theta_{h-1} \sim  \mbox{Inv-Ga}(a_2,\theta_{h-1})$. Exploiting this result 
\begin{eqnarray}
f_{\theta_{h+1} \mid \theta_{h-1}}(\theta)&=&\int_0^{+\infty}f_{\theta_{h+1} \mid\theta_h= x}(\theta)f_{\theta_h \mid \theta_{h-1}}(x) {\rm d}x= \frac{\theta_{h-1}^{a_2}}{\Gamma(a_2)^2}\theta^{-a_2-1}\int_0^{+\infty} x^{-1}e^{-\frac{\theta_{h-1}}{x}-\frac{x}{\theta}}{\rm d}x \nonumber \\
&=& \frac{\theta_{h-1}^{a_2}}{\Gamma(a_2)^2}\theta^{-a_2-1}\int_0^{+\infty} y^{-1}e^{-y-\frac{\theta_{h-1}/\theta}{y}}{\rm d}y,
\label{eq6}
\end{eqnarray}
where the last equality in \eqref{eq6} follows after the change of variable $x/\theta=y$. To evaluate the integral in \eqref{eq6}, note that from the theory of Bessel functions $\int_0^{+\infty} y^{-(\nu+1)} \exp(-y-z^2/4y) {\rm d}y=2(z/2)^{-\nu}K_{\nu}(z)$, where $K_{\nu}(\cdot)$ is the modified Bessel function of the second kind with parameter $\nu$; refer for example to \citet{wats_1966} page 183. Exploiting this result and noticing that  $\theta_{h-1}/\theta=(2\sqrt{\theta_{h-1}/\theta})^2/4$, equation  \eqref{eq6}  can be rewritten as
\begin{eqnarray}
f_{\theta_{h+1} \mid \theta_{h-1}}(\theta)=\frac{2\theta_{h-1}^{a_2}}{\Gamma(a_2)^2}\theta^{-a_2-1} K_{0}(2\sqrt{\theta_{h-1}/\theta}). 
\label{eq7}
\end{eqnarray}
Once $f_{\theta_{h+1} \mid \theta_{h-1}}(\theta)$ is available as in \eqref{eq7} let us study $\lim_{\theta \to 0^{+}} \{f_{\theta_{h+1} \mid \theta_{h-1}}(\theta)/f_{\theta_{h} \mid \theta_{h-1}}(\theta) \}$. In particular
\begin{eqnarray*}
\lim_{\theta \to 0^{+}} \{f_{\theta_{h+1} \mid \theta_{h-1}}(\theta)/f_{\theta_{h} \mid \theta_{h-1}}(\theta) \}&=&\lim_{\theta \to 0^{+}} \left[\frac{2\theta_{h-1}^{a_2}}{\Gamma(a_2)^2}\theta^{-a_2-1} K_{0}(2\sqrt{\theta_{h-1}/\theta}) \frac{\Gamma(a_2)}{\theta_{h-1}^{a_2}}\theta^{a_2+1} e^{\frac{\theta_{h-1}}{\theta}}\right] \\
&=&\lim_{\theta \to 0^{+}}  \left[\frac{2}{\Gamma(a_2)}e^{\frac{\theta_{h-1}}{\theta}+\log\{K_{0}(2\sqrt{\theta_{h-1}/\theta})\}} \right] \\
&=&\lim_{\theta \to 0^{+}} \left[\frac{2}{\Gamma(a_2)}e^{\frac{\theta_{h-1}}{\theta}-2\sqrt{\frac{{\theta_{h-1}}}{{\theta}}}-\frac{1}{4}\log\left(\frac{\theta_{h-1}}{\theta}\right)+\log\left(\frac{\sqrt{\pi}}{2}\right)}\right]\\
&=&\lim_{\theta \to 0^{+}} \left[\frac{2}{\Gamma(a_2)} e^{\frac{\theta_{h-1}}{\theta}\left\{1-2\sqrt{\frac{{\theta}}{{\theta_{h-1}}}}+\frac{\theta}{4\theta_{h-1}}\log\left(\frac{\theta}{\theta_{h-1}}\right)+\frac{\theta}{\theta_{h-1}}\log\left(\frac{\sqrt{\pi}}{2}\right) \right\}}\right]>1,
\end{eqnarray*}
for every $\theta_{h-1}>0$, $a_2>0$ and $h=2, \ldots, k-1$, where the third equality follows from results in page 378 of \citet{abra_1964}  ensuring that $K_{\nu}(z) \approx \sqrt{\pi/(2z)}e^{-z}$ for every $\nu \in \Re$  as  $z \to +\infty$. Note that in our case $z=2\sqrt{\theta_{h-1}/\theta}  \to +\infty$, when $\theta \to 0^+$. Finally, since $f_{\theta_{h+1} \mid \theta_{h-1}}(\theta)$ is an upper bound of  $f_{\theta_{h} \mid \theta_{h-1}}(\theta)$ for every $\theta_{h-1}>0$ as $\theta \to 0^{+}$, it easily follows that  $f_{\theta_{h+1}}(\theta)=\mbox{E}_{\theta_{h-1}}\{f_{\theta_{h+1} \mid \theta_{h-1}}(\theta) \}$ is an upper bound of  $f_{\theta_{h}}(\theta)=\mbox{E}_{\theta_{h-1}}\{f_{\theta_{h} \mid \theta_{h-1}}(\theta) \}$ for every $a_2>0$, proving Lemma \ref{teo2} for $h=2, \ldots, k-1$. 

When $h=1$, the proof proceeds in a similar manner after noticing that $\theta_1= \vartheta_{1}\sim  \mbox{Inv-Ga}(a_1,1)$ and $\theta_2 \mid \theta_1 \sim \mbox{Inv-Ga}(a_2,\theta_1)$. Therefore, adapting derivations in equations \eqref{eq6}--\eqref{eq7} to this case 
\begin{eqnarray*}
f_{\theta_{2}}(\theta) &=&\int_0^{+\infty}f_{\theta_{2} \mid\theta_1= x}(\theta)f_{\theta_1}(x) {\rm d}x= \frac{\theta^{-a_2-1}}{\Gamma(a_2)\Gamma(a_1)}\int_0^{+\infty} x^{-(a_1-a_2+1)}e^{-\frac{1}{x}-\frac{x}{\theta}}{\rm d}x \nonumber \\
&=&  \frac{\theta^{-a_2-1}\theta^{a_2-a_1}}{\Gamma(a_2)\Gamma(a_1)}\int_0^{+\infty} y^{-(a_1-a_2+1)}e^{-y-\frac{(2\sqrt{1/\theta})^2}{4y}}{\rm d}y= \frac{2\theta^{-a_1-1}\theta^{\frac{a_1-a_2}{2}}}{\Gamma(a_2)\Gamma(a_1)}K_{a_1-a_2}(2\sqrt{1/\theta}).
\end{eqnarray*} 
Hence, in this case the limit inequality of interest is
\begin{eqnarray*}
\lim_{\theta \to 0^{+}} \{f_{\theta_{2}}(\theta)/f_{\theta_{1} }(\theta) \}&=&\lim_{\theta \to 0^{+}} \left[ \frac{2\theta^{-a_1-1}\theta^{\frac{a_1-a_2}{2}}}{\Gamma(a_2)\Gamma(a_1)}K_{a_1-a_2}(2\sqrt{1/\theta}) \Gamma(a_1)\theta^{a_1+1} e^{\frac{1}{\theta}}\right] \\
&=&\lim_{\theta \to 0^{+}}  \left[\frac{2}{\Gamma(a_2)}e^{\frac{1}{\theta}+\log\{K_{a_1-a_2}(2\sqrt{1/\theta})\}+\frac{a_1-a_2}{2} \log(\theta)} \right] \\
&=&\lim_{\theta \to 0^{+}} \left[\frac{2}{\Gamma(a_2)} e^{\frac{1}{\theta}\left\{1-2\sqrt{\theta}+\frac{\theta}{4}\log(\theta)+\theta\log\left(\frac{\sqrt{\pi}}{2}\right)+\frac{a_1-a_2}{2}\theta \log(\theta) \right\}}\right]>1,
\end{eqnarray*}
for every $a_1>0$ and $a_2>0$, concluding the proof of Lemma \ref{teo2}.
\qed
\end{pf}

Lemma  \ref{teo2} is appealing in guaranteeing that the mass assigned by the prior to small neighborhoods of zero increases with $h$, for all $a_1>0$ and $a_2>0$, facilitating growing shrinkage. Note also that for this property to hold, the condition $a_2>1$ --- suggested by \citet{bha_2011} --- is not necessary. However, these neighborhoods might be substantially small for practical interest. Hence, from an applied perspective, it is worth assessing whether $\mbox{pr}\{\theta_1 \in (0, \theta] \} \leq \cdots \leq \mbox{pr}\{\theta_h \in (0, \theta] \}\leq  \mbox{pr}\{\theta_{h+1} \in (0, \theta] \} \leq \cdots \leq \mbox{pr}\{\theta_k \in (0, \theta] \}$  holds for a larger set of values $\theta>0$ --- and not just as $\theta \to 0^+$. This requires studying the solutions  of the inequality  $F_{\theta_{h+1}}(\theta)-F_{\theta_{h}}(\theta) \geq 0$, for each $h=1, \ldots, k-1$. As previously discussed, derivation of the cumulative distribution function for the product of independent inverse gammas  is a cumbersome task. Few  results are obtained for the product of two gammas \citep{wit_2013}. However, also in these simpler settings, analytical forms are available only for specific values of the parameters via combinations of modified Bessel and Struve functions. 

To overcome this issue and provide researchers with simple strategies to check how specific hyperparameter choices affect the cumulative shrinkage property of the prior, let us exploit the Markovian structure of the multiplicative inverse gamma prior which guarantees that $\theta_h \mid \theta_{h-1}$ is independent to $\theta_{h-2}, \ldots, \theta_1$ and $\theta_h \mid \theta_{h-1} \sim \mbox{Inv-Ga}(a_2,\theta_{h-1})$ for each $h=2, \ldots, k$. Exploiting this property it is possible to write the  inequality $F_{\theta_{h+1}}(\theta)-F_{\theta_{h}}(\theta) \geq 0$ as $\mbox{E}_{\theta_{h}}\{F_{\theta_{h+1} \mid \theta_{h}}(\theta)\}-\mbox{E}_{\theta_{h-1}}\{F_{\theta_{h} \mid \theta_{h-1}}(\theta)\} \geq 0$. Since the expectation $\mbox{E}_{\theta_{h}}\{F_{\theta_{h+1} \mid \theta_{h}}(\theta)\}$ requires the probability density function for a product of inverse gammas, and provided that this quantity is available via transformations of Meijer G-functions, the quantities $\mbox{E}_{\theta_{h}}\{F_{\theta_{h+1} \mid \theta_{h}}(\theta)\}$ are not analytically available in a simple form for $h=2, \ldots, k-1$. Hence, to address our goal let us focus on finding the solutions for the numerical approximation  of $\mbox{E}_{\theta_{h}}\{F_{\theta_{h+1} \mid \theta_{h}}(\theta)\}-\mbox{E}_{\theta_{h-1}}\{F_{\theta_{h} \mid \theta_{h-1}}(\theta)\} \geq 0$, which can be easily obtained as
\begin{eqnarray}
\mbox{E}_{\theta_{1}}\{F_{\theta_{2} \mid \theta_{1}}(\theta)\}-F_{\theta_{1} }(\theta)&\approx& \frac{1}{N}\sum_{r=1}^{N}\frac{\Gamma(a_2,\theta^{(r)}_1/\theta)}{\Gamma(a_2)}-\frac{\Gamma(a_1,1/\theta)}{\Gamma(a_1)} \geq 0, \quad \quad  \quad \quad \quad  \ \  h=1,\label{approx1}\\
\mbox{E}_{\theta_{h}}\{F_{\theta_{h+1} \mid \theta_{h}}(\theta)\}-\mbox{E}_{\theta_{h-1}}\{F_{\theta_{h} \mid \theta_{h-1}}(\theta)\} &\approx& \frac{1}{N}\sum_{r=1}^{N} \frac{\Gamma(a_2,\theta^{(r)}_h/\theta)}{\Gamma(a_2)}-\frac{1}{N}\sum_{r=1}^{N} \frac{\Gamma(a_2,\theta^{(r)}_{h-1}/\theta)}{\Gamma(a_2)} \geq 0,  \quad h \geq 2, \ \ \ \
\label{approx2}
\end{eqnarray}
where generic samples $\theta^{(r)}_h$ are easily available as cumulative products of $h$ independent inverse gammas from \eqref{eq4}.

In order to provide guidelines for possible behaviors of the multiplicative gamma process prior, Table \ref{table_fin} reports solutions of $F_{\theta_{h+1}}(\theta)-F_{\theta_{h}}(\theta) \geq 0$ approximated via \eqref{approx1}--\eqref{approx2} --- for each $h=1, \ldots, k-1$ --- focusing on different combinations of $a_1>0$ and $a_2>0$, with $k=5$. According to our discussion, the intersection of these solutions --- computed for each $h=1, \ldots, k-1$ --- provides the  set of values $\theta>0$  for which $\mbox{pr}\{\theta_1 \in (0, \theta] \} \leq  \mbox{pr}\{\theta_2 \in (0, \theta] \}\leq  \mbox{pr}\{\theta_{3} \in (0, \theta] \}  \leq \mbox{pr}\{\theta_4 \in (0, \theta] \} \leq \mbox{pr}\{\theta_5 \in (0, \theta] \}$ holds. For example, when $a_1=2$ and $a_2=2$, $\mbox{pr}\{\theta_1 \in (0, \theta] \} \leq  \mbox{pr}\{\theta_2 \in (0, \theta] \}\leq  \mbox{pr}\{\theta_{3} \in (0, \theta] \}  \leq \mbox{pr}\{\theta_4 \in (0, \theta] \} \leq \mbox{pr}\{\theta_5 \in (0, \theta] \}$ is guaranteed for all intervals $(0, \theta]$ such that $\theta\leq \bar{\theta} \approx 1.79$.

\begin{table}
\begin{center}
\def~{\hphantom{0}}
{%
\begin{tabular}{lcccccc}
&& $h=1 \rightarrow h=2$ & $h=2 \rightarrow h=3$  & $h=3 \rightarrow h=4$  &$h=4 \rightarrow h=5$ &Intersection\\
\hline
 ~$(a_1=1,a_2=1)$ && $(0 ,\ \approx 0{.}52]$ & $(0,\ \approx0{.}33]$ & $(0,\ \approx0{.}22]$  &$(0,\ \approx0{.}14]$&$(0,\ \bar{\theta}\approx0{.}14]$\\
 ~$(a_1=1,a_2=2)$ && $(0,\ >100]$ &$(0,\ >100]$ & $(0,\ >100]$&$(0,\ >100]$&$(0,\ \bar{\theta}>100]$\\
 ~$(a_1=1,a_2=3)$ && $(0,\ >100]$ & $(0,\ >100]$ & $(0,\ >100]$ &$(0,\ >100]$&$(0,\ \bar{\theta}>100]$\\
 \hline
 ~$(a_1=2,a_2=1)$ && $(0,\ \approx0{.}33]$ & $(0,\ \approx0{.}21]$ & $(0,\ \approx0{.}13]$  &$(0,\ \approx0{.}09]$&$(0,\ \bar{\theta}\approx0{.}09]$\\
 ~$(a_1=2,a_2=2)$ && $(0,\ \approx1{.}79]$ &$(0,\ \approx3{.}18]$ & $(0,\ \approx5{.}67]$&$(0,\approx9{.}90]$&$(0,\ \bar{\theta}\approx1.79]$\\
 ~$(a_1=2,a_2=3)$ && $(0,\ >100]$ & $(0,\ >100]$ & $(0,\ >100]$ &$(0,\ >100]$&$(0,\ \bar{\theta}>100]$\\
 \hline
\end{tabular}}
\caption{Solutions of \eqref{approx1}--\eqref{approx2} for $k=5$, based on different combinations of $a_1$ and $a_2$. In evaluating \eqref{approx1}--\eqref{approx2}, $N=1{,}000{,}000$. For every combination of $a_1$ and $a_2$, the intersection of the solutions comprises the set of values $\theta>0$ for which \eqref{approx1}--\eqref{approx2}  holds for every $h=1, \ldots, k-1$}
\label{table_fin}
\end{center}
\end{table}

For the combinations of $a_1$ and $a_2$ considered, the value of $\bar{\theta}$ increases as $a_2$ grows, for every $a_1>0$. An opposite behavior is instead observed when $a_1$ increases, for all $a_2>0$. Moreover, it is worth noticing how $\bar{\theta}$ becomes substantially large when $a_2$ is moderately higher than $a_1$. When instead $a_1=2$ and $a_2=1$, the prior assigns increasing mass only to intervals $(0, \theta]$, with $\theta\leq \bar{\theta} \approx 0{.}09$. Note also how --- according to Table \ref{table_fin} --- the width of the intervals for which $\mbox{pr}\{\theta_{h+1} \in (0, \theta] \}  \geq \mbox{pr}\{\theta_{h} \in (0, \theta] \}$ is true, decreases for growing $h$, when $a_2=1$. In fact, $\bar{\theta}$ may become even smaller than $0{.}09$, when  $k > 5$. Hence, although cumulative shrinkage holds for any $a_1>0$ and $a_2>0$ in small neighborhoods of zero, based on the above considerations, avoiding excessively high values for $a_1>0$ and setting $a_2 \geq 2$ and moderately higher than $a_1>0$, can provide a better choice to facilitate shrinkage. 

Although  Table \ref{table_fin} focuses on few standard cases, the proposed procedures provide the researchers with the basic guidelines and simple strategies to study the stochastic behavior of the multiplicative inverse gamma prior at all possible combinations of $a_1>0$, $a_2>0$ and $k$. Note also that, when studying the first two moments of the induced prior on the elements in $\Omega$, \citet{bha_2011} require $a_1>2$ and $a_2>3$ in order to ensure the existence of these prior moments. Although motivated by substantially different probabilistic properties, this restriction is also a good setting to induce appropriate shrinkage based on the above analyses.

\section{Simulation study}
In assessing the shrinkage effects induced by the multiplicative gamma process on posterior inference, let us focus on the factor model proposed in   \citet{bha_2011}  and outlined in Section \ref{sec1}. In particular, the aim of this simulation is to empirically evaluate how different hypeparameter choices in the multiplicative inverse gamma prior affects posterior inference on covariance matrices $\Omega^0$ factorized as $\Omega^0=\Lambda^0 \Lambda^{0\T} + \Sigma^{0}$, at varying rank $k^0$. In accomplishing this goal let us focus on the dependence structure among $p=10$ Gaussian random variables --- simulated for $n=100$ units --- with $\Sigma^0=\mbox{I}_p$ and $\Lambda^{0}$ a $p \times k^0$ matrix with entries sampled from standard Gaussian random variables. In a first simulation scenario $k^0=2$, whereas in the second $k^0=6$.

\begin{figure}[t]
\centering
\includegraphics[trim=0.7cm 0.77cm 0.1cm 0.1cm, clip=true, width=12.74cm]{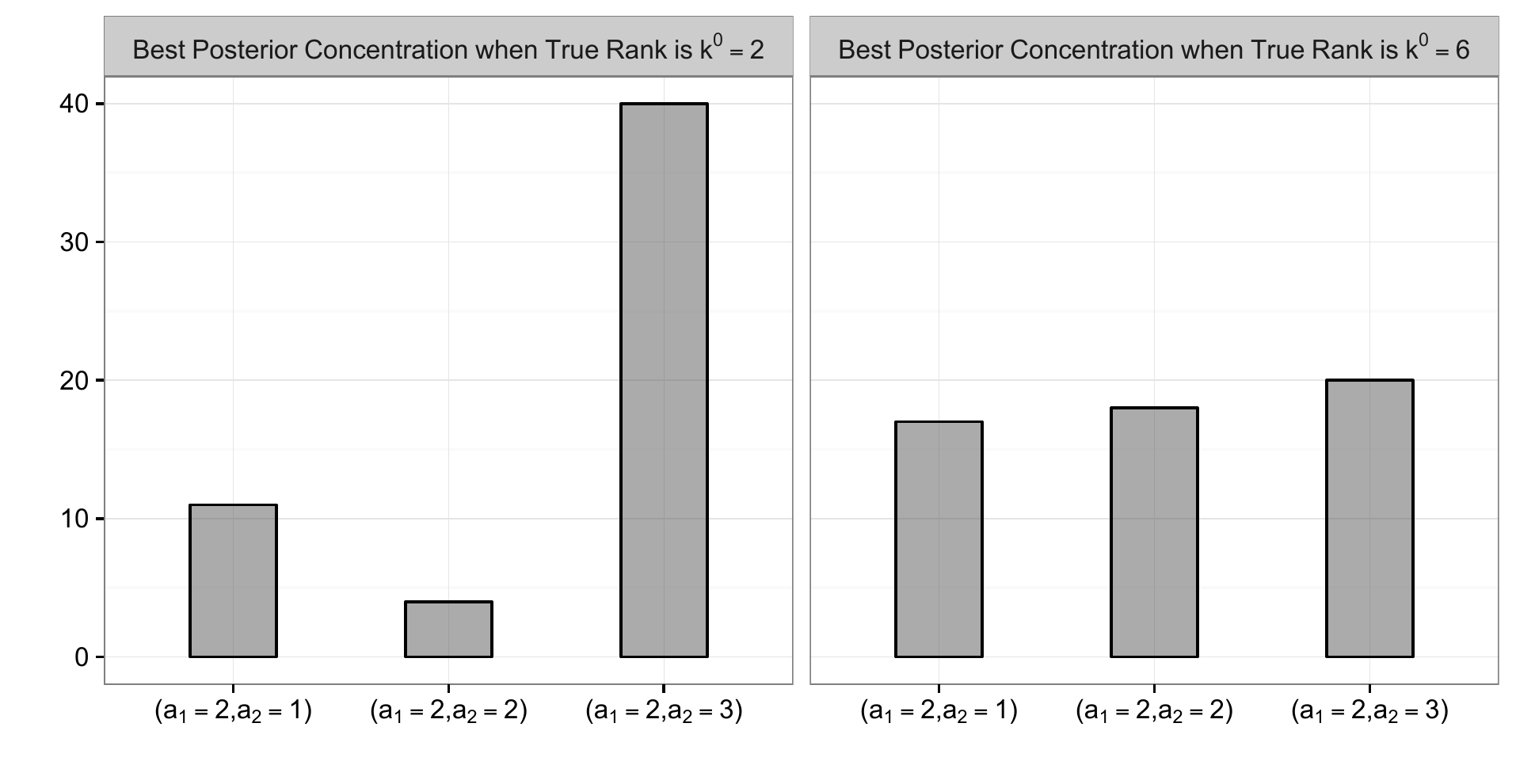}
\caption{Total number of unique parameters in $\Omega$ for which each prior setting induces the best posterior concentration, compared to the others.}
\label{f1}
\end{figure}

In both simulation scenarios, posterior computation is performed leveraging the steps of the Gibbs sampler proposed in  \citet{bha_2011}, with the same settings as in their simulation and fixing the highest possible rank $k=10$ as a truncation level. In order to empirically assess the type of shrinkage induced on the posterior distribution of the parameters in $\Omega$, inference is performed for three different choices of the hyperparameters in the multiplicative inverse gamma prior, including $(a_1=2, a_2=1)$, $(a_1=2, a_2=2)$ and $(a_1=2, a_2=3)$. Figure \ref{f1} compares the performance of posterior inference for the selected hyperparameters and $k^0$, by showing the total number of different parameters $\omega_{js}$, $j=1, \ldots, p$, $s \leq j$ in $\Omega$ for which each prior setting  induces the best posterior concentration, compared to the others. Posterior concentration, for each $\omega_{js}$, $j=1, \ldots, p$, $s \leq j$, is measured via $d_{js}=\mbox{E}\{(\omega_{js}-\omega^0_{js})^2 \mid y_1, \ldots, y_n\}$, with this expected squared deviation from the truth estimated from $30{,}000$ Gibbs samples, after a burn-in of $5{,}000$. These MCMC settings were sufficient for convergence and good mixing.

As shown in Figure \ref{f1} the multiplicative gamma process with hyperparameters  $a_1=2$ and $a_2=3$ induces posterior distributions for the parameters $\omega_{js}$, $j=1, \ldots, p$, $s \leq j$  which are typically characterized by an improved posterior concentration. The performance gains are more evident when the true rank is $k^0=2$, than $k^0=6$. These results are consistent with Table \ref{table_fin}, showing that the order in probabilities holds for wider intervals when $a_1=2$ and $a_2=3$, compared to $(a_1=2, a_2=1)$ and $(a_1=2, a_2=2)$. This result allows stronger shrinkage, facilitating improved posterior concentration towards lower-dimensional factorizations. However, when the true rank increases, the induced multiplicative shrinkage might be too strong. This motivates further research to define prior distributions characterized by adaptive shrinkage, which do not excessively penalize higher dimensional factorizations.

Although the setting  $(a_1=2, a_2=3)$ induces improved posterior concentration for a wider subset of parameters $\omega_{js}$, $j=1, \ldots, p$, $s \leq j$, the overall performance of the three hyperparameters choices remains on comparable values. In fact, when $k^0=2$, the median of the estimated squared deviations ${d}_{js}$, $j=1, \ldots, p$, $s \leq j$ is $0.067$, $0.066$ and $0.065$ for $(a_1=2, a_2=1)$, $(a_1=2, a_2=2)$ and $(a_1=2, a_2=3)$, respectively. When $k^0=6$, these quantities are instead $0.371$, $0.353$ and $0.357$. This result is consistent with Lemma \ref{teo2}, showing that the order in probabilities holds --- at least in small intervals of zero --- for any choice of the hyperparameters characterizing the multiplicative inverse gamma prior. In fact, all the three settings improve posterior concentration compared to a situation in which the prior distribution induces no shrinkage. Specifically, the median of the estimated squared deviations ${d}_{js}$, $j=1, \ldots, p$, $s \leq j$ is $0.081$ for $k^0=2$ and $0.375$ for $k^0=6$, when Bayesian inference is performed by replacing the multiplicative gamma prior with $\tau_{h} \sim \mbox{Ga}(2,2)$, for each $h=1, \ldots, k$.

\section{Discussion}
The aim of this contribution has been on providing novel probabilistic insights on the multiplicative gamma process. As discussed in the note, this prior is characterized by tractable computational strategies and induces flexible shrinkage, making it an appealing candidate in several applications, and worth a deeper probabilistic understanding. The main focus has been on assessing the shrinkage properties motivating the multiplicative gamma process construction and how they relate to its hyperparameters. The contribution refines initial results in  \citet{bha_2011} and provides strategies to study the shrinkage behavior at varying $(a_1, a_2)$ and $k$. A simulation study exhibits how the shrinkage properties associated with the prior are translated to posterior inference. Beside these practical motivations, an improved  understanding of such process can facilitate future studies on the  theoretical performance of the posterior distribution in recovering the true model dimensions. Early results are available in simple models \citep{rou_2011}, and it is an active area of research to extend these properties to more general scenarios. 

\section*{Acknowledgment}
\vspace{-4pt}
I would like to thank David Dunson, Anirban Bhattacharya, Willem van den Boom, Bruno Scarpa and Antonio Canale for the helpful comments on a first version of this note.

\vspace{-4pt}

\bibliography{paper-ref}


\end{document}